\documentclass[runningheads]{llncs}
\usepackage{graphicx}
\usepackage{changepage}
\usepackage{subcaption}
\begin{document}
\title{SSLEM: A Simplifier for MBA Expressions based 
on Semi-linear MBA Expressions and Program Synthesis
}

\titlerunning{ SSLEM: A Simplifier for MBA Exprs. based 
on Semi-linear MBA Exprs. and P. Synthesis}
%
 \author{
  Seong-Kyun Mok\inst{1} \and
  Seoyeon Kang\inst{1} \and
  Jeongwoo Kim\inst{1} \and
  Eun-Sun Cho\inst{1}  \and
  Seokwoo Choi\inst{2}}
 \authorrunning{S.-K. Mok et al.}
%
 \institute{Chungnam National University, Daejeon, South Korea \and
 The Affiliated Institute of ETRI, Daejeon, South Korea\\
 \email{sy.kang@o.cnu.ac.kr, mok7764@cnu.ac.kr, bnm961126@gmail.com, eschough@cnu.ac.kr, seogu.choi@gmail.com}}	
\maketitle              
\begin{abstract}
MBA (mixed boolean and arithmetic) expressions are hard to simplify, so used for malware obfuscation to hinder analysts' diagnosis. 
	Some MBA simplification methods with high performance have been developed, 
	but they narrowed the target to "linear" MBA expressions, 
	which allows efficient solutions based on logic/term-rewriting. 
	However such restrictions are not appropriate for general forms of MBA expressions usually appearing in malware. 
	To overcome this limitation, we introduce a "semi-linear" MBA expression, 
	a new class of MBA expression extended from a linear MBA expression, 
	and propose a new MBA simplifier called "SSLEM", based on a simplification idea of semi-linear MBA expressions and program synthesis. 
\end{abstract}

\section{Introduction}
MBA (Mixed Boolean and Arithmetic) expressions are 
the expressions mixed with arithemethic operators like `$+$', `$-$' and $\times$',
and boolean operators like `$\&$', `$|$' and `$\wedge$'.
Program rewriting to insert MBA expressions is a popular way of program obfuscation
because such expressions are difficult to analyze with popular solvers like Z3 and Mathematica.

While MBA obfuscation is a helpful method for software protection,
it is also used for malware obfuscation. 
Once malware is  obfuscated with the MBA-obfuscation option, it entails a herculean effort of analysts to understand its behavior.
Thus there have been various simplification efforts to handle this problem.

The methods used by existing works to simplify MBA expressions are classified into two kinds; 
First, logic-based simplifiers have been devised to simplify MBA expressions.
Based on theoretical foundation \cite{Eyrolles17}\cite{Liu21},
this logic-based approach suffers from the scalability problem \cite{Zhou07}. 
Thus most of the previous works narrowed the problem scope.
For instance, typically logic-based simplifiers have been dedicated to a specific form of MBA expressions like linear MBA expressions
to make the problem tractable, systematic and fast \cite{Liu21}.
However, it does not seem appropriate for obfuscated malware, 
because trivial MBA expressions shown in obfuscated malware like $x \times y$, $x \ \& \ ( x + y)$ and  $ x \ \& \ 0x1$  are not linear expressions.

Second, synthesis-based MBA expression simplication \cite{Blazytco17}\cite{David20}\cite{David21} has been recently introduced.  
This approach searches a simpler expression of the original program, if any, by program synthesis.
It is applicable to general complicated forms of MBA expressions, besides specific forms of expressions.
However, the solution is imcomplete and relatively slow \cite{David21}.

In this poster, we proposed a method and tool to develop an enhanced simplifier of MBA expressions
to help deobfuscate MBA-obfuscated malware, taking advantage of program synthesis as well as logic-based solutions.

\section{Previous Works} 
Previous works recognized nested categories of MBA expressions.
``Linear MBA expressions'' is a summation of terms, 
where a summation  means an arithmetic operation with $+$, 
a term is a multiplication of a coefficient $a_i$ and a pure boolean expression $e_i(x_1, \dots x_T)$,
$T$ is the number of variables 
and $N$ is the number of terms, 
as depicted in the following fomula \cite{Eyrolles17}\cite{Liu21}\cite{Zhou07};

\begin{equation}
	\sum_{i=0}^{N} a_i e_i (x_1, \dots x_T) \label{eqn:linear}
\end{equation}

Examples of linear expressions include $10\times(x \ \& \ y)$, $5\times(x \ | \ y) + 6\times(x \ \& \ y)$, $3 \times (x \ \& \ y) - 20 \times (x  \ | \ y) -1$ and $ y + 376 \times (x \ \& \ -1)$.

As mentioned earlier, most traditional simplification of MBA expressions dedicated to linear MBA expressions.
Depending on pattern matching and rule rewritting, 
they are not feasible for complicated non-linear expressions due to the lack of the scalability.

It is proved a valuable proposition that 
a simplified form of any of the linear MBA expressions with 1-bit variables also represents 
a simplified form of the one with the n-bit variables, and vice versa \cite{Liu21}\cite{Eyrolles17}.
This means that if we simplify any linear MBA expression with 1-bit variables, 
the simplification result can be also considered as a simplification of the corresonding, same looking expression with n-bit variables \cite{Liu21}.
This makes the simplification process much more efficient.

For some insight, let us consider an expression a function which maps a combination of input variables to the output.
Regardless representation of an expression whether it is complicated or simple,
the number of unique mappings will be limited by the size and the number of the input and output variables.
For instance,
if we use two 1-bit variables $x$ and $y$ as inputs, the possible combination of $x$ and $y$ will be 00, 01, 10 and 11.
The simplification of a given linear MBA expression would be 
a searching problem. 
Among all the possible output-bit-combinations for each input-bit-combination,
a simplifier should find a simpler expression corresponding to the given expression.
The property proven by the previous works dramatically reduced the searching space from 
$2^{2^{2^{16}}}$ (that is, $16^{16}=1.84e+19$)  to $2^{2^{2^1}}$ (that is, $8$) .

Note that the proposition proved by previous methods only for linear MBA expressions.
However, solution dedicated to linear MBA expressions cannot handle a lot of general MBA expressions-only $2^{2^{2^1}}$ expressions can be handled.
Thus, they adopted heuristics to spot any linear subexpressions in a given complicated expression.
Otherwise $2^{2^{2^16}} - 2^{2^{2^1}}$ MBA expressions cannot be handled, 
some of which might appear in obfuscated malware.

\section{SSLEM: The Proposed MBA Expression Simplifier}
In this section, we introduce SSLEM, a new MBA expression simplification solution 
To overcome limitations of existing works, SSLEM makes use of both kinds of existing tools  
(the logic-based simplifier and the program synthesis-based simplifier),
we envision that the proposed mehod will enhance performance some corner cases.

As mentioned earlier, a logic-based simplifier is fast and effiecient 
but simplifying only limited category of expressions (linear MBA expressions).
Aiming to overcome this problem, the proposed SSLEM proposes to 
preprocess the general MBA expressions to linear MBA expressions,
and input the result to a logic-based  simplifier.
However, transformation to linear expressions would fail in most cases,
because linear MBA expressions are only a small subset of the whole MBA expressions, 
as we compared the number of unique semantics in the previous section.

Thus, SSLEM uses an intermediate form of MBA expressions called ``semi-linear'' MBA expressions, 
and transforms a general MBA expressions to a semi-linear MBA expression, first, by the program synthesis-based simplifier.
Then the resulting semi-linear MBA expression is transformed into a combined expression which is made of
several linear expressions.
Since each of the resulting linear MBA expressions can be simplified by the logic-based simplifier, 
we can finally earn simplification of a non-linear expression with a logic-based simplier.

	\begin{figure}[!t]
	\includegraphics[width=130mm]{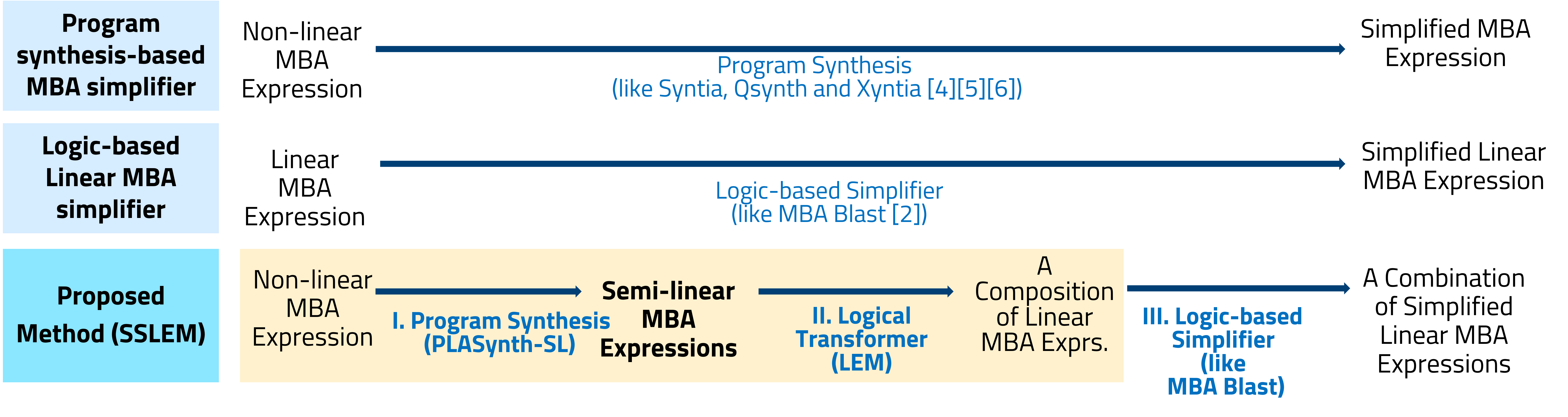}
	\caption{ Overview of the proposed method and existing method}
    	\label{fig:overall}
\end{figure}

Overall process of the proposed SSLEM is composed of three steps, as depicted in Fig.~\ref{fig:overall}:

\begin{enumerate}
	\item PLASynth-SL finds preliminary candidates 
		for the logic-based simplifier working at Step III.
		It is a search-based synthesizer to find a semi-linear MBA expression 
		corresponding to a general non-linear MBA expressions, keeping the semantics the same.

	\item LEM transforms a semi-linear MBA expression 
		to a combination of linear MBA expressions

	\item The linear subexpressions of the semi-linear MBA expression resulting from LEM, 
		are in turns fed to the logic-based simplifier.
		This step might leverage an existing powerful linear MBA expression simplifier like MBA Blast \cite{Liu21}

\end{enumerate}

\subsection{Semi-linear MBA expressions and PLASynth-SL}
Let us start with a ``heterogeneous-bit-valued constant'',  a constant that has different bit values in it  (eg. 0xFE,  0x1, 0x003F),
Note that a linear expression only a ``homogeneous-bit-valued constant'' in its boolean subexpressions, which includes 0xFF, -1 and 0x0, 
	each of which consist of the same valued bits (all 1's or all 0's).
Semi-linear MBA expressions are in a slightly extended form of linear expressions, defined as follows; 
\begin{definition}
A semi-linear MBA expression is an MBA Expression that could be a linear MBA expression, 
		if without ``heterogeneous-bit-valued constants'' 
		in the boolean subexpressions.
		Semi-linear expressions inlucde "$30 \times (0xFE \ | \ y)$'', ``$21 \times x \ \& \ 0x01)$''
		
\end{definition}

PLASynth-SL is equipped with optimization and timeout management \cite{Tigress} 
for better performance.
In a preliminary experiment, 
out of the 500 MBA expression samples obfuscated by Tigress \cite{Xu21} with the MBA-obfuscation option, 
PLASynth-SL earned 68 semi-linear MBA expressions, 
while gained 46 linear MBA expressions.

\subsection{LEM (Linear MBA Expression Generator from Semi-linear MBA Expressions)}

\begin{figure}[!t]
		\centering
	\includegraphics[width=55mm]{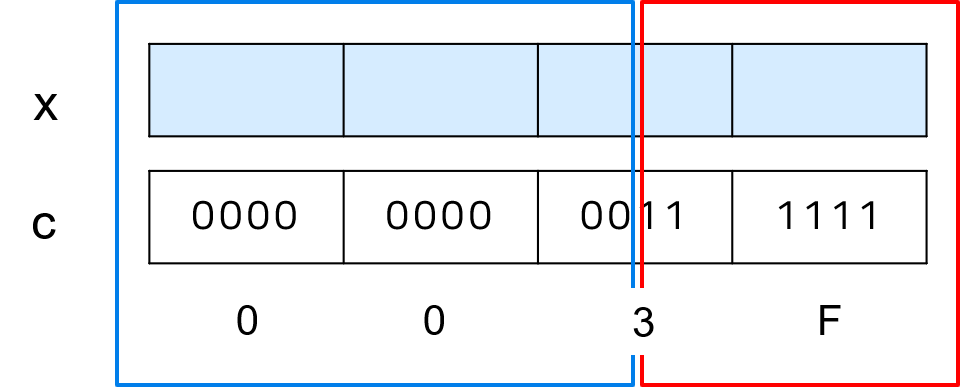}
		\caption{ A bit representation of a heterogeneous-bit-valued contstant and the variable}
	\label{fig3}
    \label{fig:bitsplit}
\end{figure}

Let us assume that a given 16-bit semi-linear MBA expression $f$ is  ``$ x \wedge 0x003F$'',
Here, ``0x003F'' is the heterogenous-bit-valued constant, without it, $f$ could be a linear MBA expression.
As depicted in Fig.~\ref{fig:bitsplit}, this constant is divided into 0000 0000 00  and 11 1111.

The MBA expression $f$ is considered as a combination of $f_1$ and  $f_2$,
where $x$ = [$X_1^{10},  X_0^{6}$] in the multi-granularity representation \cite{Xu18}
$f_1$ = $X_1^{10} \wedge 0x00000$, and
$f_2$ = $X_0^{6} \wedge 0x3F$.
At this time,  each of $f_1$ and $f_2$  is a linear MBA expression. 
This means that
the simplified expression of $f_1$, say $F_1$, is syntatically same to $F_1^{'}$, 
where $F_1^{'}$ is the simplified expression of $f_1$ when interpreted 
as a 1-bit expression \cite{Eyrolles17}\cite{Liu21}\cite{Zhou07}.
(Similarly, $F_2$, is syntatically same to $F_2^{'}$.) 
as a 1-bit expression. 
Thus, the final result of the simplification from $f$ is formulated as $(f_1 << 6) + f_2$.

\section{Discussions and Conclusions}
We propose 
a novel MBA simplifier, called ``SSLEM,'' based on a new concept of ``semi-linear'' MBA expression transformation, 
as well as existing logic based simplifiers that restrict their target  into linear MBA expressions.
We assume that more semi-linear MBA expressions exist in real-world obfuscated malware, than linear MBA expressions do.
The shorter contiguous sequences of the same bits in a -bit-valued constant complicates the result. 
For instance, 0101 0101 0101 01 0101 will require 16 separated linear expressions and 
15 shifts in the simplification result, which might not be simpler than the original expression.
However, our method will be effective with the small number of changes in the bits of constants like 0x01 and 0xFE.
By taking the advantages of program synthesis and logically proved solutions, we envision that SSELM will show better performance.
Since some expressions are likely to be simplified by program synthesis, while some others are simplified by logic-based rewriting methods, 
we expect that SSLEM would be able to handle the corner cases that the existing solutions could not hanldle.

\end{document}